# Graphical abstract

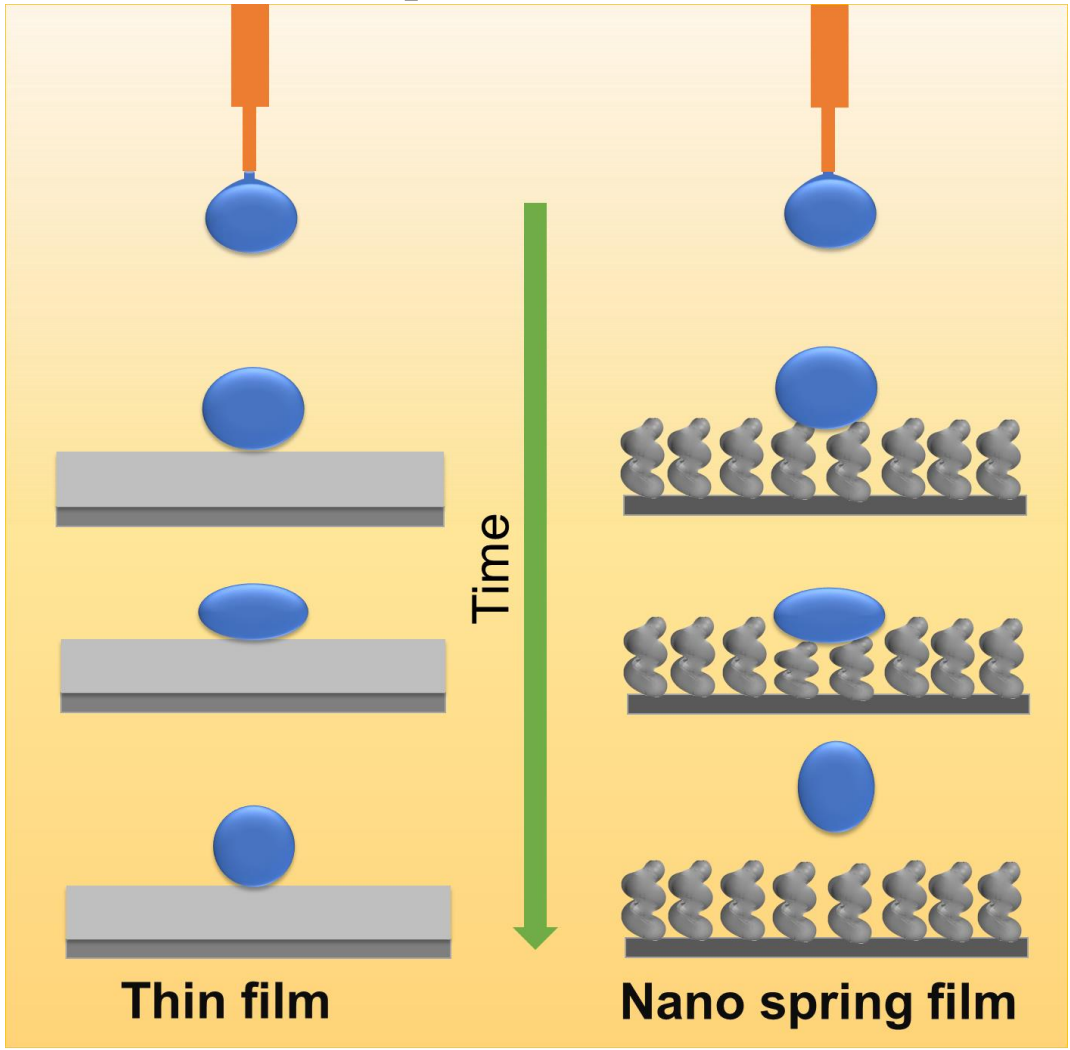

# Water Droplet Bouncing on a non-Superhydrophobic Si Nanosprings

*Samir Kumar,[1,2] Kyoko Namura,[1] Motofumi Suzuki,[1] and Jitendra P. Singh[2,*]*

[1]Department of Micro Engineering, Graduate School of Engineering, Kyoto University, Nishikyo, Kyoto 615-8540, JAPAN.

[2]Department of Physics, Indian Institute of Technology Delhi, Hauz Khas, New Delhi 110016, INDIA.

## ABSTRACT

Self-cleaning surfaces often make use of superhydrophobic coatings that repel water. Here, we report a hydrophobic Si nanospring surface, that effectively suppresses wetting by repelling water droplets. We investigated the dynamic response of Si nanospring arrays fabricated by glancing angle deposition. The vertical standing nanospring arrays were approximately 250 nm tall and 60 nm apart, which allowed the droplets to rebound within a few milliseconds after contact. Amazingly, the morphology of the nanostructures influences the impact dynamics. The rebound time and coefficient of restitution were also found to be higher for Si nanosprings than vertical SI columns. It has been proposed that the restoring force of the Si nanosprings may be responsible for the water droplet rebound and can be explained by considering the droplet/nanospring surface as a coupled spring system. These nanospring surfaces may find applications in self-cleaning windows, liquid-repellent exteriors, glass panels of solar cells, and antifouling agents for roof tiling.

## INTRODUCTION

Micro and nanostructured surfaces with special wetting behaviors have received considerable attention in recent years because of their use as self-cleaning surfaces.[1–6] Non-wettability is a crucial surface property that plays an important role in daily life, industry, and agriculture. The lotus effect is an example of self-cleaning in nature, where superhydrophobic leaves protect the lotus plant against pathogens or fungi. [7] Depending on the surface energy and ruggedness of its microstructures, a surface can be hydrophilic, hydrophobic, or superhydrophobic.[8] Superhydrophobic surfaces can be fabricated by chemically modifying a surface with a low surface energy coating and by creating a surface from a hydrophobic material that exhibits roughness at the micro- or nanoscale.[9] For any practical applications, the superhydrophobicity and non-wetting behavior must be maintained under dynamic conditions that are when the droplet impacts the surface with a certain velocity. On superhydrophobic surfaces, water will form almost spherical droplets with very high contact angles. When landing on such a surface, the water droplet may rebound, which is critical for situations where the impact of water droplets on the surface is encountered, for example, in deicing applications.[10,11] There are several reports on various necessary conditions for the bouncing. First, bouncing can be easily achieved on superhydrophobic surfaces as there is little interaction between the droplet and surface, which might otherwise prevent the drop from bouncing.[12] When a droplet falls on such a surface, the rough structures of the surface and the air trapped by the droplet can offer a significant capillary pressure to help the droplet rebound of the surface.[13–16] Several studies have elucidated the ensuing dynamics of a bouncing droplet[16–22] as a function of the surface micro- and nanostructure,[18,23,24], and as a function of the impact velocity.[25] The shape-change in the droplet has also been shown to be a direct indicator of the contact angle and hydrophobicity. Bouncing of the water droplets have been studied as a parameter to determine the hydrophobicity of the surface, and a relationship has been established between the contact angle of the water and the number of bounces.[26] It has also been reported that the surface must have a contact angle of at least 151° for a droplet to bounce so that the kinetic energy of the impinging droplet can be transferred to the surface energy.[12,26–28] Second, there are studies that suggest that the hysteresis of the contact angle plays a crucial role in the bouncing behavior of the impacting droplets.[29] Apart from the wetting property of the surface, the rebound also depends on other parameters like surface tension, viscosity, and velocity of drop at impact.[18,20,22,25,27]

There are many reports on water droplet bouncing on superhydrophobic surfaces.[22,30–32] Bouncing droplets are generally reported for high contact angle (superhydrophobic) static surfaces, but bouncing on a hydrophobic nanospring structure, to the best of our knowledge, has not been reported previously. Little work has been reported about the droplet rebound on hydrophilic or hydrophobic surfaces.[33,34] Here, we show that an ultrathin film of nanosprings can cause water droplets to rebound. We demonstrate that the nanostructured surfaces from the same material that has comparable static contact angles exhibit remarkably different droplet rebound dynamics. Even though millions of nanostructures interact simultaneously with a single water droplet, the underlying shape of the nanostructures can determine the direction in which the droplet flies-off the surface.

## MATERIALS AND METHODS

i.  **Samples: fabrication of nanostructures**

We used the glancing angle deposition (GLAD) technique to fabricate nanostructured surfaces with different morphologies.[35,36] GLAD can be used to grow porous and hydrophobic surfaces from a variety of materials[37–43]. Depositions were performed with vacuum chamber pressures in the range $10^{-7} – 10^{-6}$ torr by electron-beam evaporation. The angle between the substrate normal and the incident vapor flux was 85° during the entire deposition. The substrate was rotated continuously and slowly for the fabrication of Si nanosprings arrays.[44,45] We prepared thin film, slanted column, vertical column, and nanospring arrays of Si for this study. Henceforth, these samples will be referred to as TF, SC, VC, and SS, respectively. The samples were coated with trichlorooctadecylsilane to modify the sticking behavior of the surface.

**ii. SEM analysis**

The as-grown Si nanostructures were imaged using a scanning electron microscope (SEM; Hitachi High Tech. SU3800) with a $LaB_6$ detector in the secondary electron mode operating at an acceleration voltage of 10 kV. The cross-sectional SEM images were acquired by cleaving the wafer to expose a pristine edge.

**iii. Drop impact experiments**

For the measurement of the static apparent contact angle (APCA), 10 $\mu L$ droplets of ultrapure water (18.2 MΩ cm from Millipore Direct Q UV3, Merck) of about 1.3 mm radius were gently dropped onto the substrate. The contact angle measurements were repeatedly performed at ten different positions on the sample surfaces. The impact and rebound dynamics of the water droplet were followed by using a high-speed camera (FASTCAM Mini AX100, Photron) operating at 3000 frames/s.[46] A 10 μL volume of water droplets was used for the dynamic measurements, and the droplets were allowed to fall due to gravity on the nanostructured sample surfaces. The impact velocity ($v_0$) was changed by varying the droplet release height ($h$). Water droplets were positioned at 10, 15, and 20 mm above the surface with corresponding impact velocities of 44, 54, and 63 cm s$^{-1}$ and Weber number ($W_e$) of 7.2, 10.8, and 14.3, respectively. The impact velocity of the water droplet was calculated using the relation $v_0 = \sqrt{2gh}$, where $g$ is the acceleration due to gravity. All measurements were performed at room temperature and 37% relative humidity.

## RESULTS AND DISCUSSION

Figure 1 shows the top and cross-sectional SEM micrographs of the different Si nanostructures used in this study. The inset shows optical images of 10 μL volumes of water droplets on the corresponding surfaces. Average thickness, average diameter, and solid fraction of each sample are given in Table 1.

**Table 1.** Average thickness, average diameter, and solid fraction of the samples

| Sample | Average thickness (nm) | Average diameter (nm) | Solid fraction (%) | |
|---|---|---|---|---|
| | | | from contact angle | from SEM image |
| TF | 553±2 | NA | NA | NA |
| SC | 265±8 | 53±15 | 35 | 43 |
| VC | 240±2 | 48±15 | 20 | 27 |
| SS | 256±4 | 45±10 | 35 | 45 |

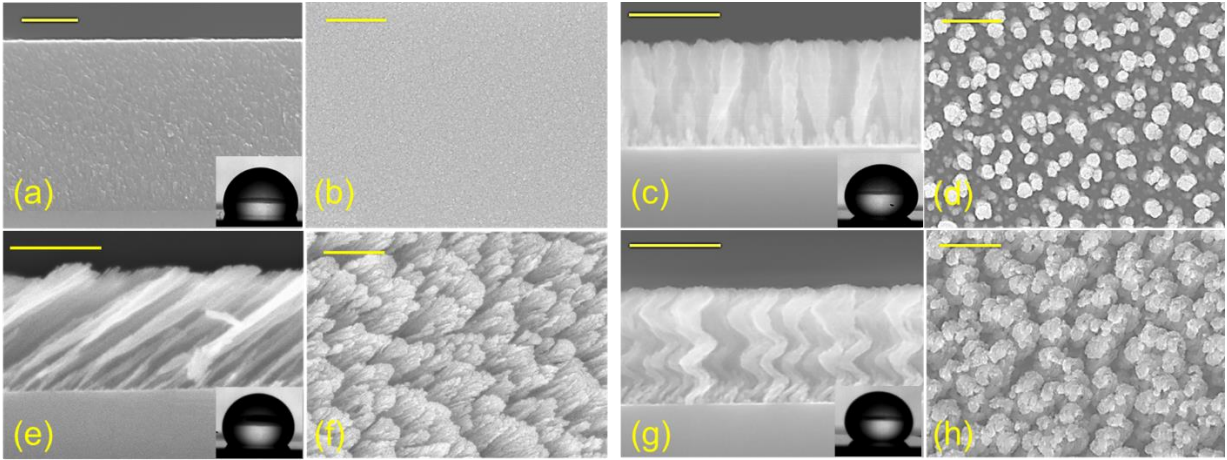

**Figure 1.** Top and cross-sectional SEM images of thin film(a-b); vertically standing nanorod(c-d); slanted nanorods (e-f); and Si nanospring arrays (g-h), respectively.

All the nanostructured samples had a thickness of approximately 250 nm and an average diameter of approximately 50 nm. Generally, when a droplet is placed statically on a periodic nanostructured surface, the droplet shape is symmetric and is determined by the minimization of the total surface energy. The static APCA values of water droplets on TF, SC, VC, and SS were observed to be 106°, 135.7°, 148.6°, and 138.6°, respectively, Fig. 2. The contact angle on the TF was minimum and was maximum for the SS. The SC and SS had a similar contact angle. However, the water droplet impact dynamics were found to be very different.

The chemical composition of the surface and surface morphology defines the wetting property of a surface. All the samples were made of Si and were coated with the same chemical (which resulted in a slightly higher contact angle); hence, the reason for the difference in contact angle is the surface morphology of the samples. The nanocolumnar structure makes the sample surface very rough and porous, resulting in an increase in contact angle as compared to the conventional thin film. The contact angle was found to increase from 106° for the conventional film to 148° for the vertical nanocolumnar sample. The increase in contact angle on the nanocolumnar samples can be attributed to the decrease in the solid fraction of the nanostructures, as per the Cassie-Baxter model.[47] The solid fraction $f$ is given as

$$f = \frac{\cos \theta_A + 1}{\cos \theta_0 + 1} \qquad (1)$$

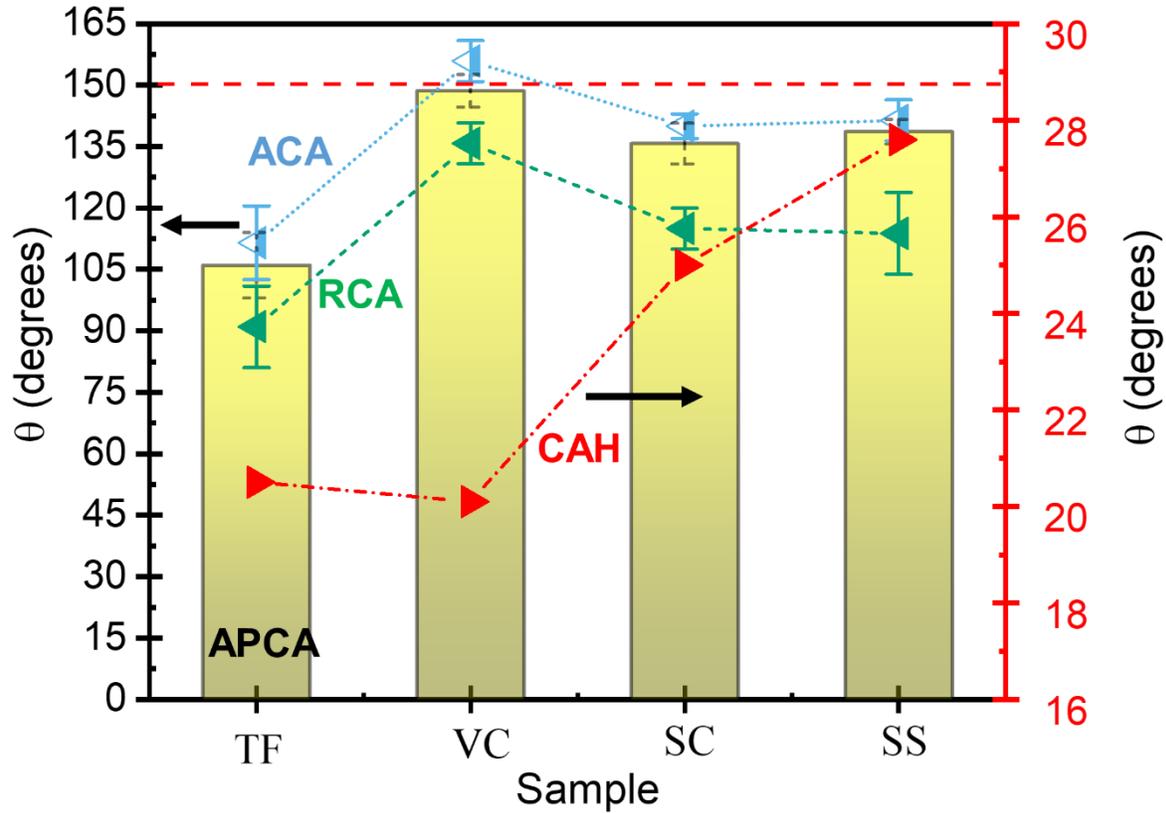

**Figure 2.** The static apparent contact angle, APCA, (bar graph), advancing contact angle, ACA, (blue curve), receding contact angle, RCA (green curve) and contact angle hysteresis, CAH (red curve) values of water droplets on TF, SC, VC, and SS.

where $\theta_A$ is the apparent contact angle on the nanostructured surface, and $\theta_0$ is the contact angle on a conventional surface. In the Cassie-Baxter state surface is a composite of air and Si, and the water droplet sits on the air trapped between the rough surfaces with apparent contact angle $\theta_A$. The calculated solid fraction using Eq. 1 and the SEM images are in good agreement. The slightly lower solid fraction calculated from the Cassie–Baxter model may because of the change in contact angle due to the chemical modification. A similar increase in contact angle with the nanocolumnar structure has also been reported in previous studies.[48–50] The spreading dynamics of a water droplet on the vertical Si nanocolumns have also been studied by other groups, but we are interested only in the bouncing behavior of water droplets on these nanostructures.[51]

The droplet size, liquid viscosity $\mu$, and impact velocity $v_{0s}$ all influence the impact dynamics. A dimensionless variable Weber number $W_e$ (the ratio of the kinetic energy to the surface energy) can be used to characterize the impact dynamics[52]

$$W_e = \frac{\rho D_0 v_0^2}{\sigma} \quad (2)$$

where $D_0$ is the droplet diameter, $\rho$ is the density, and $\sigma$ is the surface tension of the liquid. In this paper, $\rho$, $D_0$, and $\sigma$ are fixed, so we modified $W_e$ only by changing the $v_0$. Droplets can rebound for high $W_e \geq 10$.[53] For this purpose, we performed experiments at $W_e \approx 7$, 10, and 14. Water droplets of 10 µL volume (diameter ≈ 2.67 mm) were dropped onto the silanized Si nanostructured samples. Water droplets were released from a height of 10 mm above the surface with corresponding impact velocities of 44 cm s$^{-1}$ and $W_e$ of 7.2. When the droplet was dropped from any height on the slanted nanorods, the droplet deformed to an ellipsoidal shape and then recoiled without detaching from the surface (Fig. 3).

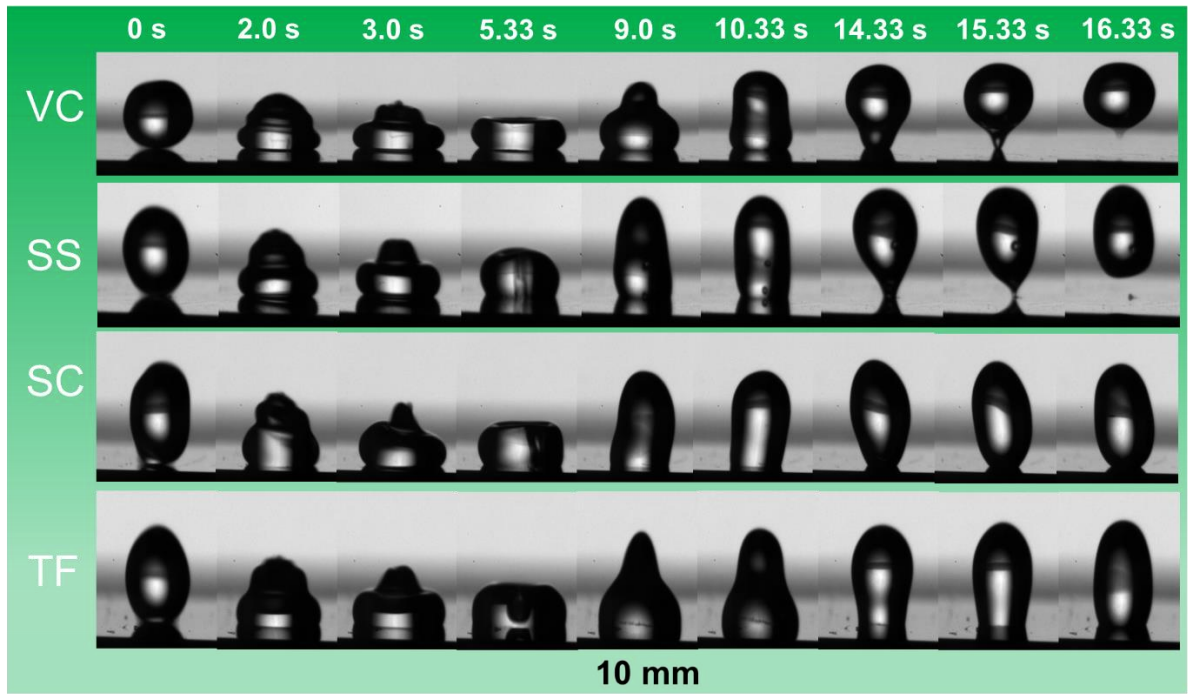

**Figure 3.** Time evolution of 10 µL water droplets dropped from a height of 10 mm on different nanostructures.

A similar phenomenon was observed on thin-film for all impact velocities. The maximum spreading diameter ($d_{max}$) depends on the impinging velocity of the droplet, the capillary and viscous forces, as well as the properties of the liquid and the solid surface.[54] When evaluating the spreading behavior of a droplet, the maximum spreading diameter is usually normalized with respect to its initial diameter ($d_0$) as the dimensionless spreading factor, $\beta_m = d_{max}/d_0$. The maximum spreading factors for TF, SC, VC, and SS were approximately 1.47, 1.36, 1.39, and 1.36, respectively. The $\beta_m$ was found to increase with increasing contact angle and also with increasing $W_e$, Fig 4. The increase in the $\beta_m$ with an increase in the contact angle can be explained using the equation[55]

$$\beta_m = \sqrt{\frac{W_e + 12}{3(1 - \cos\theta_a) + 4\left(W_e / \sqrt{R_e}\right)}} \quad (3)$$

where $\beta_m$ is the spreading factor, $\theta_a$ is the advancing contact angle, $W_e$ is the Weber number, and $R_e$ is the Reynolds number. The spreading mechanism of a drop onto a solid surface has been studied in detail in the past.[56] The evolution of the spreading factor is divided into four phases: the kinematic, spreading, relaxation, and wetting phases, respectively. Most of the spreading occurs during the spreading phase, which is dominated by inertia.[57] The increase in inertia can explain the increase in the maximum spreading diameter with increasing $W_e$.

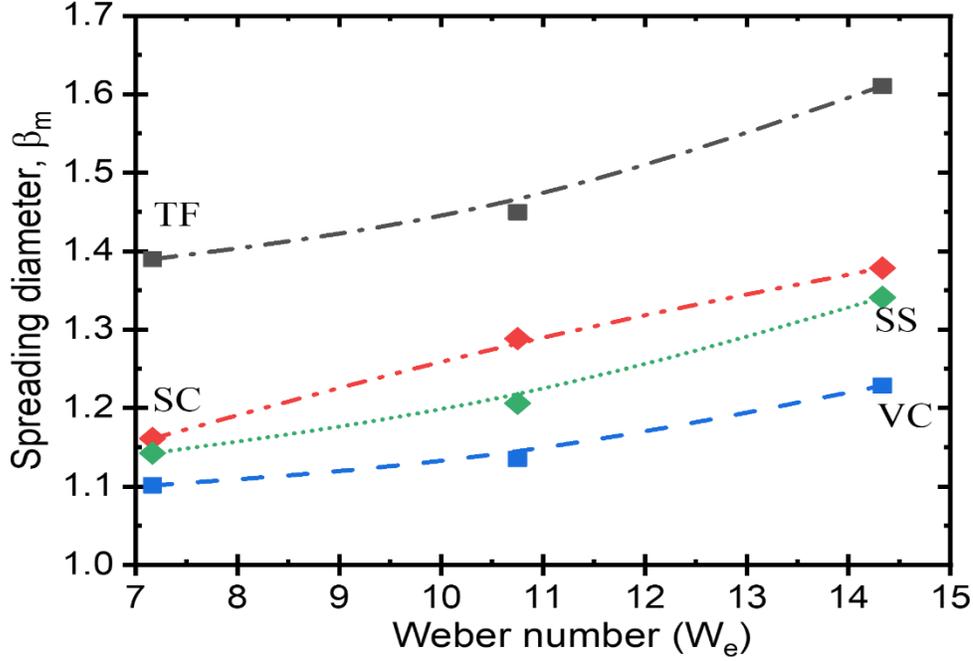

**Figure 4.** The dimensionless spreading factor, $\beta_m = d_{max}/d_0$ vs $W_e$ for TF, SC, VC, and SS. The $\beta_m$ was found to increase with increasing contact angle and also with increasing $W_e$.

**Table 2.** *Velocity, Weber number, maximum deformation, time for maximum deformation, rebound time, and coefficient of restitution for the four samples. The time to reach maximum deformation was also found to depend on the impacting surface and was maximum for TF and minimum for SS sample. .*

| Sample | Velocity (cm/s) | Weber no. $W_e$ | maximum spreading diameter ($d_{max}$) (mm) | Time for max. deformation (ms) | Rebound Time (ms) | Time of flight (ms) | Coefficient of restitution (COR) |
|---|---|---|---|---|---|---|---|
| TF | 44.3 | 7.2 | 3.92 | 4.67 | - | | |
| | 54.2 | 10.8 | 4.04 | 4.11 | - | | |
| | 62.6 | 14.3 | 4.47 | 3.56 | - | | |
| SC | 44.3 | 7.2 | 3.72 | 4.89 | - | | |
| | 54.2 | 10.8 | 3.9 | 3.56 | -- | | |
| | 62.6 | 14.3 | 4.1 | 3.33 | - | | |
| VC | 44.3 | 7.2 | 3.63 | 4.11 | 16.11 | 22.67 | 0.25 |
| | 54.2 | 10.8 | 3.69 | 3.56 | 15.67 | 27.22 | 0.25 |
| | 62.6 | 14.3 | 3.92 | 3.33 | 16.33 | 29 | 0.23 |
| SS | 44.3 | 7.2 | 3.7 | 3.87 | 15.33 | 14.68 | 0.16 |
| | 54.2 | 10.8 | 3.82 | 3.67 | 17.42 | 19.33 | 0.18 |
| | 62.6 | 14.3 | 4.12 | 3.44 | 18.58 | 26.33 | 0.21 |

The time to reach maximum deformation was also found to depend on the impacting surface and was maximum for TF and minimum for SS, Table 2. Sample SS took around 3.67 ms, on average, for maximum deformation for $We \approx 7$. The droplet reached its maximum deformation at approximately $t$ = 4.11 ms for the VC sample, after which surface tension and viscous forces overcame inertia, so that fluid accumulated at the leading edge of the splash and started pulling back. Droplets with higher velocity will have higher inertia and will take less time for maximum deformation. Hence, the time for maximum deformation decreased with an increase in impact velocity.

The rebound time at which the droplet bounces off the surface is crucial because it determines the degree of energy transfer. When the droplet falls on vertically aligned nanorods from a height of 10 mm, it rebounded and left the surface in ≈ 16.11 ms. However, when the same volume of droplets was freed with the same impact velocity of 44 cm/s on the Si nanosprings (APCA < $150^0$), then instead of wetting the surface, the droplet bounced off and left the surface in ≈ 15.33 ms. The SS structure not only showed the bouncing of the droplet on the hydrophobic surface but also reduced the contact time (≈15.33) and the time for maximum spreading (≈ 3.67 ms) than that of VC samples. [32] It is interesting to note that the rebound time for the VC sample was almost constant (≈ 16 ms) with increasing $W_e$, but the rebound time for the SS sample increased from 15.3 to 18.5 ms. The spreading dynamics, in the case of VC, is consistent with the previous report by Fan et al.[51]

The bouncing behavior on the VC is not unexpected because it has a contact angle of ≈ 148.6° ± 4. In other words, it satisfied the first necessary condition for bouncing behavior. The contact angle for the SS is around 138.6° ± 3, but surprisingly it also shows the bouncing behavior. Some reports concluded that contact angle hysteresis (CAH) plays a significant role in the bouncing from the surface.[29] The process of impact of a droplet is an interplay between the kinetic energy, surface energy, and viscosity of the water droplet. The elastic force is due to the surface tension of the water droplet, and viscosity is the cause of energy dissipation. Before the impact with the surface, the droplet possesses only kinetic energy. When the drop impacts the solid, the drop gets deformed, and a shock wave spreads radially outward towards the surface up to the point when the viscosity dissipates kinetic energy. The dissipation due to heat can be neglected as it is infinitesimal for water. When the droplet reached its maximum deformation, the restitution force due to surface tension comes into play, which causes the droplet to recoil. Now the droplet shrinks and moves radially inward and gains kinetic energy, and a jet rises in the center (Worthington jet), which may lead to the lift of the droplet (Fig. 3). The droplet must do some work to overcome the resistance force produced by the CAH. The total work done, $W$ in the spreading and receding process of a droplet is given by[29]

$$W = \frac{1}{8}\pi\beta_m^2 D_0^2 \gamma_{LV}(\cos\theta_r - \cos\theta_a) \tag{4}$$

where $\cos\theta_r$ and $\cos\theta_a$ are the receding and advancing contact angles, $(\cos\theta_r - \cos\theta_a)$ is the CAH, and $D_0$ is the initial diameter of the droplet. As a consequence, lower CAH values may result in lower work against the resistive force and will require very little energy to overcome the work done, resulting in a rebound. The lower CAH value for the VC sample may be one of the reasons for the rebound despite having a longer contact time. SC and SS have similar CAH values of around 25 and 27.6, but only SS showed the rebounding property.

On superhydrophobic surfaces, the dynamics of a droplet impinging on a surface depend on the competition between the three wetting pressures: water hammer pressure, $P_{wh} = \rho C_w v_0$, dynamic pressure, $P_d = \rho \frac{v_0^2}{2}$, and the anti-wetting capillary pressure, $P_c = -2\sqrt{2}\gamma_{LV}\cos\theta_a/B$ where $\rho$ is the water density, $C_w$ is the speed of sound in water, $V_i$ is the droplet velocity, $\gamma_{LV}$ is the surface energy of the water at the water and vapor interface, $\theta_A$ is the advancing contact angle, and $B$ is the spacing between the nanostructures.[58,59] Capillary pressure is caused by the air trapped by the surface roughness. The air cushion trapped between the nanorods and the water droplet acts as an effective spring. For a droplet to rebound from the surface, the non-wetting condition $P_c > P_{wh} > P_d$ has to be satisfied. The water hammer pressure $P_{wh}$ and dynamic pressure $P_d$ was found to vary from 0.66 - 0.93 MPa (considering $\rho$ = 1000 kg m$^{-3}$ and $C_w$ = 1482 m s$^{-1}$) and 0.1 – 0.2 kPa respectively, for the three experimental heights in increasing order. The capillary pressure $P_c$ was calculated as 2.44 MPa, 2.76 MPa, and 4.33 MPa for the SC, SS, and VC, respectively. The capillary pressure generated by the VC was maximum. For SS and SC nanostructured surfaces, $P_c$ was also comparable to each other; however, the bouncing phenomenon was observed only on the nanosprings surface. Thus, the capillary pressure alone cannot be the only reason for the observed bouncing behavior of the droplet on the nanospring surface.

The rebound of the droplet on the surface of the nanosprings is surprising, as it is generally assumed that only superhydrophobic surfaces support bouncing, as the capillary pressure forces on the superhydrophobic surfaces are such that they can allow a drop from the surface. A detailed model for the rebound of the water droplet on vertically aligned nanorods can be found in the supplementary information. Hence, we propose the hypothesis that the elastic property of the nanospring has a significant role in the bouncing of the water droplet. The rebound of a droplet is only possible if the kinetic energy of the impinging droplet is larger than the surface energy dissipated during the retraction stage. Bouncing water droplets are vertically deformable, and upon impact, some of the kinetic energy can be stored by the deformation of the droplet itself.[60] The droplet itself thus behaves like a spring, whose stiffness is the surface tension.[28] The nanosprings can store sufficient energy to facilitate a rebound that causes the droplet to detach from the surface completely.

We modeled the elasticity of the droplet in contact with the elastic nanosprings as an effective mechanically coupled double-spring system. To understand this effect, we propose to model the droplet by two identical masses *m* linked by a spring of stiffness $k_w$ and a rest length *L*. The viscous effects into the spring are modeled by a mechanical damper with dissipation parameter *β*. The coordinate *y* is taken vertically upward, and the vertical positions are $y_1$ for the upper mass and $y_2$ for the lower mass. The schematic representation of the spring on a spring system is given in Fig. 5. The force of gravity acting on the two masses is $F_{g1}$ = -mg = $F_{g2}$ . The spring also exerts forces on each mass given by $F_{s1}$ = -$k_w(y_1 – y_2 – L)$ and $F_{s2}$ = $k_w( y_1 – y_2 – L)$. When the lower mass is in contact with the nanospring, it experiences a normal force $F_{ns}$ = - $k_{ns}y_3$, where $k_{ns}$ is the stiffness constant of the nanospring. The normal force $F_{ns}$ depends on the compression of the nanospring, which wary during contact. This normal force is zero when the droplet is not in contact with the surface. The motion of both the masses can be described in the laboratory frame by the following set of coupled Newton's equations in the laboratory fr

$$m\frac{d^2 y_1}{dt^2} + \beta(\frac{dy_1}{dt} - \frac{dy_2}{dt}) + k_w(y_1 - y_2 - L) + mg = 0 \qquad (6)$$

$$m\frac{d^2 y_2}{dt^2} - \beta(\frac{dy_1}{dt} - \frac{dy_2}{dt}) - k_w(y_1 - y_2 - L) + mg = -k_{ns} y_3$$

$$m\frac{d^2 y_3}{dt^2} = -k_{ns} y_3$$

The constant $k_w$ models the undamped frequency of the spring given by $f_0 = \sqrt{(k_w/m)}$.

The loss of energy of two objects after a collision can be described in terms of the coefficient of restitution (COR), that is, COR depends on the elastic properties of the colliding objects. Since, in this study, one of the colliding objects was always water droplet, the COR will depend on the elastic property of the nanostructured surface. COR was found to be almost constant for VC and increase for SS with an increase in $W_e$. A decrease in the COR value of VC for $W_e$ 14 may be due to the fact that the air trapped under the water droplet may be forced out because of the higher impact velocity of the droplet.

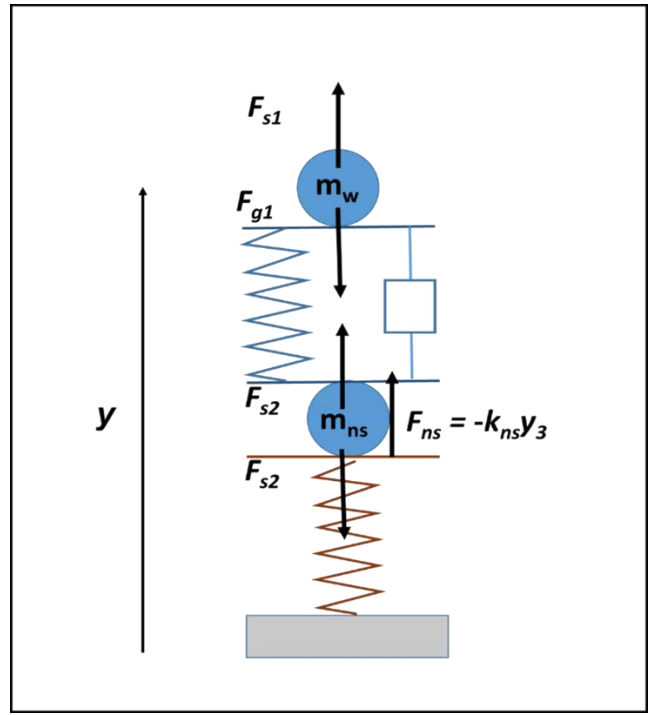

**Figure 5.** Representation of a spring on a spring bouncing system considering the water droplet as an elastic spring.

We can explain the increase in the rebound time and COR for the SS sample if we consider the compression of the nanospring structure on droplet impact. The relation between the initial velocity $v_0$, maximum compression $y_m$ of the nanospring, and spring constant is given by[61]

$$y_m^2 = v_0^2 \frac{m}{k} \qquad (7)$$

A higher velocity will lead to a higher compression, which may increase the rebound time on the nanospring sample. The potential energy of the nanosprings is also directly proportional to the square of the maximum compression and spring constant. The nanospring will absorb higher energy for higher $W_e$ and return a higher fraction of energy on recoil. Haneko et al., have already studied the elastic behavior of a Si nanospring fabricated using the GLAD technique.[62] They showed that Si nanospring exhibit nonlinear elastic mechanical behavior. They reported the load–displacement (F–δ) relationships obtained during the loading and unloading processes. The nanospring showed a nonlinear reversible behavior, and the relationship between the load F [nN], and displacement, δ [nm], was determined to be F = 4.1δ + 0.0041δ2. They also confirmed that this nonlinearity originated from the large deformation permitted by the spring shape. Therefore, if we consider the droplet/nanospring surface as a coupled spring device, we can understand the bouncing behavior on the SS sample along with the increase in the rebound time and COR.

## CONCLUSION

The dynamics of water droplets falling on vertical nanorods, tilted nanorods, and nanosprings of silicon were studied. After impact with the surface, the water droplet initially spreads and flattens, and then recoils, which is greatly influenced by the underlying morphology of the nanostructured surface. On slanted Si nanorods, no recoil was observed, whereas, on vertical and nano helices, the recoil was completed in approximately 16 ms. Interestingly, water droplets were observed to bounce on hydrophobic nanosprings with higher rebound time and COR than that of vertical nanostructures. The elastic force arising from the difference between the equilibrium droplet shape and the deformed droplet shape drives the recoiling flow. The restoring force of the nanosprings may be responsible for the rebound of the water droplet and can be explained by considering the droplet/nanospring surface as a coupled spring system.

**ACKNOWLEDGMENTS**

**ASSOCIATED CONTENT**

**Supporting Information**

The rebound of the water droplet on vertically aligned nanorods.

## Acknowledgments

The authors also thank Dr. Kosuke Ishikawa of Kyoto University for assisting us with the SEM observations.

## Funding

This work was supported by the grant Center of Innovation Program (COI) from the Japan Science and Technology Agency (JST), "The Last 5X Innovation R&D Center for a Smart, Happy, and Resilient Society" (grant number JPMJCE1307).

## Competing Interests

The authors declare that there is no conflict of interest.

## AUTHOR INFORMATION

## Corresponding Authors

*E-mail: jpsingh@physics.iitd.ac.in, samiratwork@gmail.com